\documentclass[10pt,a4paper]{article}
\usepackage{jheppub_kim}
\usepackage{pdflscape}
\usepackage{amsmath}
\usepackage{amssymb}
\usepackage{dcolumn}
\usepackage{bm}
\usepackage{color}
\usepackage{epsfig}
\usepackage{amsfonts}
\usepackage{graphicx}
\usepackage{subfigure}
\usepackage{dcolumn}
\usepackage{hyperref} 

\begin{document}
\title{ Quantum gravitational corrections to the geometry of charged
AdS black holes}
\author[a,b,c,d]{Behnam Pourhassan,}
\author[e]{Ruben Campos Delgado,}
    \author[f,g,a]{Sudhaker Upadhyay,\footnote{Visiting Associate, Inter-University Centre for Astronomy and Astrophysics (IUCAA) Pune-411007, Maharashtra, India}}
    \author[a]{Hoda Farahani,} 
    	         \author[a]{and Himanshu Kumar}

\affiliation[a]{School of Physics, Damghan University, P. O. Box 3671641167, Damghan, Iran }
\affiliation[b]{Center for Theoretical Physics, 
Khazar University,  41 Mehseti Street, Baku, AZ1096, Azerbaijan}
\affiliation[c]{Physics Department, Istanbul Technical University, Istanbul 34469, Turkey}
\affiliation[d]{Centre of Research Impact and Outcome, Chitkara University, Rajpura-140401, Punjab, India}
\affiliation[e]{Bethe Center for Theoretical Physics, Physikalisches Institut der Universit\"at Bonn, Nussallee 12, 53115 Bonn, Germany}
   \affiliation[f]{Department of Physics, K.L.S. College, Nawada, Magadh University, Bodh Gaya, Bihar 805110,  India}
        \affiliation[g]{Department of General \& Theoretical Physics, L. N. Gumilyov Eurasian National University,  Astana, 010008, Kazakhstan}
  
\emailAdd{b.pourhassan@du.ac.ir; b.pourhassan@candqrc.ca} 
 \emailAdd{ruben.camposdelgado@gmail.com} 
  \emailAdd{sudhakerupadhyay@gmail.com; sudhaker@associates.iucaa.in}
 \emailAdd{h.farahani@umz.ac.ir}
  \emailAdd{hkumarphysics@gmail.com}
  
 \abstract{
We consider a modified gravity by higher-derivative gravity coupled with
non-local terms and Maxwell electrodynamics. By employing the effective field theory framework for quantum gravity, we compute quantum corrections to the entropy of charged AdS black holes, focusing on second-order curvature terms. By incorporating scale-dependent coefficients, we  establish that the Wald entropy is renormalization group (RG) invariant, confirming the robustness of the framework. 
Additionally, quantum corrections to thermodynamic quantities-temperature, pressure, specific heat, and Helmholtz free energy-are derived, all satisfying the first-law of thermodynamics. Specific heat and Helmholtz free energy also exhibit RG invariance, demonstrating stability under scale transformations.
These findings highlight the effectiveness of the approach in describing quantum modifications to charged AdS black hole thermodynamics, offering insights into the interplay between quantum gravity and black hole physics.}
\maketitle
%%%%%%%%%%%%%%%%%%%%%%%%%%%%%%%%%%
%%%%%%%%%%%%%%%%%%%%%%%%%%%%%%%%%%
\section{Introduction}\label{sec:intro}
One of the most profound advancements in theoretical physics over the last half-century has been the discovery that black holes exhibit entropy \cite{PhysRevD.7.2333}, a concept that bridges the domains of thermodynamics, quantum mechanics, and general relativity. Within the framework of Einstein's general relativity, the leading-order contribution to black hole entropy is elegantly characterized as one-quarter of the area of the event horizon. Extensions or modifications to gravitational theories introduce additional terms, enriching the entropy's theoretical structure. Recently, the application of effective field theory methods \cite{PhysRevD.104.066012} has enabled a rigorous computation of quantum gravitational corrections to the entropy of a Schwarzschild black hole, leveraging the Wald entropy formalism \cite{PhysRevD.48.R3427}. 
A significant point to consider is that quantum gravitational effects can induce modifications to the metric, potentially altering the location of the event horizon. This development marks a significant step toward a deeper understanding of black hole thermodynamics in the context of quantum gravity. It is worth highlighting that black hole entropy may also receive corrections from various additional sources. For example, quantum fluctuations of matter fields in the vicinity of black hole spacetimes, along with fluctuations in the spacetime geometry within the canonical quantum gravity framework, contribute to entropy modifications. Notably, these corrections often exhibit a logarithmic dependence, reflecting their distinct origin and nature \cite{PAUL2023116259,Kumar_2023}.

The exploration of black hole thermodynamics  provides a compelling framework for testing and refining our theories of gravity \cite{PhysRevD.7.2333,PhysRevLett.26.1344,PhysRevD.9.3292}. This potential arises from the inherent simplicity of thermal systems, which can be characterized by a limited set of macroscopic parameters. In the case of black holes, these parameters are distilled into fundamental quantities such as mass, entropy, charge, angular momentum, and, depending on the gravitational model, a few additional variables.
The significance of black hole thermodynamics lies in its profound connection to the principles of quantum gravity--a field undergoing active development through promising approaches like string theory. Particularly, the quantum aspects of gravity become most apparent in the vicinity of the event horizon, where intense gravitational effects render classical thermodynamic treatments insufficient. These considerations underscore the unique role of black holes as both theoretical challenges and windows into the deeper structure of spacetime and quantum gravity.

 In this study, we employ the effective field theory framework for quantum gravity to systematically compute quantum gravitational corrections, up to second order in curvature, to the entropy of a physically significant class of black holes-namely, charged AdS black hole. Here, inserting the
explicit scale dependence of the coefficients  we find that the Wald entropy is RG invariant. Furthermore,  we derive the quantum gravitational corrections to relevant thermodynamic  quantities satisfying first-law of thermodynamics, namely, temperature, pressure, specific heat and Helmholtz free energy. Here,  we check that   both specific
heat and Helmholtz free energy are RG invariant.

The paper is structured as follows to systematically explore the quantum gravitational corrections to charged AdS black hole properties:  
   In this section \ref{sec:corrections_metric}, we delve into the modifications to the black hole metric arising from quantum gravitational corrections. We analyze how these corrections influence the structure of the black hole spacetime, particularly focusing on changes to the event horizon and other key geometric properties.  
    The section  \ref{sec:entropy} is dedicated to investigating the effects of quantum gravity on the entropy of black holes. Using the Wald entropy formalism, we compute the corrections to the classical entropy and examine their dependence on the underlying parameters of the theory, such as curvature and coupling constants.  
   Here, in section \ref{sec:thermodynamics}, we explore the quantum gravitational corrections to various thermodynamic quantities of black holes, including temperature, pressure, specific heat, and Helmholtz free energy. We demonstrate how these corrections maintain consistency with the first law of thermodynamics and assess the implications for the stability and equilibrium properties of black holes.  
In the final section, we summarize the key findings of the paper, emphasizing the implications of quantum gravitational corrections for black hole physics. Additionally, we discuss potential avenues for future research, particularly in the context of testing and validating these theoretical predictions.  

This structured approach ensures a comprehensive exploration of the subject, linking theoretical insights with broader implications for black hole thermodynamics and quantum gravity.
\section{Quantum gravitational corrections to the classical metric in AdS space}\label{sec:corrections_metric}
We start by reviewing some known facts about the effective field theory approach to quantum gravity. For a good introduction to this topic, see for example \cite{Buchbinder:1992rb, Donoghue:2017pgk}. A possible way to study quantum effects in gravity is to modify the pure Einstein-Hilbert action by including additional terms normally suppressed at low energies.   The main possibility, at second order in curvature, is the local action
\begin{equation}\label{eq:local_action}
    \Gamma_{L}=\int d^4x\, \sqrt{-g}\,\bigg(\frac{R+2\Lambda}{16\pi G_N} +c_1(\mu)R^2+c_2(\mu)R_{\mu\nu}R^{\mu\nu}+c_3(\mu)R_{\mu\nu\rho\sigma}R^{\mu\nu\rho\sigma}\bigg),
\end{equation}
where $\Lambda$ is the cosmological constant and $\mu$ is an energy scale. The exact values of the constants $c_1$, $c_2$, $c_3$ are calculable provided that one assumes an ultra-violet complete theory of quantum gravity, see for example \cite{Myrzakulov:2014hca, Elizalde:2017mrn}.
By integrating out fluctuations of the graviton and of any matter field irrelevant to the problem under consideration, one also gets a non-local (NL) effective action \cite{Weinberg:1980gg, Starobinky:1981ZhPmR, Barvinsky:1983vpp, Barvinsky:1985an, Barvinsky:1987uw, Barvinsky:1990up, Donoghue:1994dn}, which, at second order in curvature, is
\begin{equation}
    \Gamma_{NL}=-\int d^4 x \sqrt{-g}\bigg[\alpha R\ln\left(\frac{\Box}{\mu^2}\right)R+
    \beta R_{\mu\nu}\ln\left(\frac{\Box}{\mu^2}\right)R^{\mu\nu} + \gamma R_{\mu\nu\rho\sigma}\ln\left(\frac{\Box}{\mu^2}\right)R^{\mu\nu\rho\sigma}\bigg],
\end{equation}
where $\alpha$, $\beta$ and $\gamma$ are constants which can be computed in a model-independent way \cite{Donoghue:2014yha}.
The operator $\ln\left(\Box/\mu^2\right)$ has the integral representation \cite{Donoghue:2015nba}
\begin{equation}
    \ln\left(\frac{\Box}{\mu^2}\right)=\int_0^{+\infty}ds\, \left(\frac{1}{\mu^2+s}-\frac{1}{\Box+s}\right).
\end{equation}
Using both the local and non-local Gauss-Bonnet identities \cite{Calmet:2018elv}, it is possible to eliminate the Riemann tensor by redefining the coefficients as
\begin{equation}
\begin{gathered}
   c_1 \to\bar{c}_1=c_1-c_3, \hspace{2mm}c_2\to\bar{c}_2=c_2+4c_3, \hspace{2mm} c_3\to 0,\\
    \alpha\to\bar{\alpha}=\alpha-\gamma, \hspace{2mm} \beta\to\bar{\beta}=\beta+4\gamma, \hspace{2mm} \gamma\to 0.
\end{gathered}
\end{equation}
We can couple gravity to electromagnetism by adding the Maxwell action
\begin{equation}
    \Gamma_{M}=-\frac{1}{4}\int d^4x\,\sqrt{-g}\,F_{\mu\nu}F^{\mu\nu},
\end{equation}
where $F_{\mu\nu}=\partial_{\mu}A_{\nu}-\partial_{\nu}A_{\mu}$ is the electromagnetic tensor and $A_{\mu}$ is the electromagnetic potential.
In this paper we then consider the full action as
\begin{equation}\label{eq:total_action}
    \Gamma=\Gamma_{L}+\Gamma_{NL}+\Gamma_{M}\equiv\int d^4x\,\sqrt{-g}\,\mathcal{L}.
\end{equation}
The Maxwell equations, obtained by varying \eqref{eq:total_action} with respect to $A_{\mu}$, are
\begin{equation}\label{eq:maxwell_equations}
    g^{\mu\nu}\nabla_{\mu}F_{\nu\tau}=0.
\end{equation}
The quantum corrected Einstein equations, obtained by varying \eqref{eq:total_action} with respect to the metric, are
\begin{equation}\label{eq:einstein_equations}
    \frac{1}{8\pi G_N}\left(G_{\mu\nu}-\Lambda g_{\mu\nu}\right)+2\left(H_{\mu\nu}+K_{\mu\nu}\right)=T_{\mu\nu},
\end{equation}
where $T_{\mu\nu}$ and $G_{\mu\nu}$ are respectively the energy-momentum tensor and Einstein tensor, which are given by
\begin{eqnarray}
    T_{\mu\nu}&=&\frac{1}{4\pi}\left(F_{\mu\rho}{F_{\nu}}^{\rho}-\frac{1}{4}g_{\mu\nu}F_{\rho\sigma}F^{\rho\sigma}\right),\\
    G_{\mu\nu}&=&R_{\mu\nu}-\frac{1}{2}R g_{\mu\nu}.
\end{eqnarray}
However, $H_{\mu\nu}$ and $K_{\mu\nu}$ encode the quantum effects. The local contribution is
\begin{eqnarray}
    H_{\mu\nu}&=&\bar{c}_1\Big(2R R_{\mu\nu}-\frac{1}{2}g_{\mu\nu}R^2-2\nabla_{\mu}\nabla_{\nu}R+2g_{\mu\nu}\Box R\Big)\nonumber\\
   & +&\bar{c}_2 \left(-\frac{1}{2}g_{\mu\nu}R_{\rho\sigma}R^{\rho\sigma}+2R^{\rho\sigma}R_{\mu\rho\nu\sigma}-\nabla_{\mu}\nabla_{\nu}R+\Box R_{\mu\nu}+\frac{1}{2}g_{\mu\nu}\Box R\right),
\end{eqnarray}
while the non-local contribution is
\begin{eqnarray}
    K_{\mu\nu}&=& -2\bar{\alpha}\Big(R_{\mu\nu}-\frac{1}{4}g_{\mu\nu}R+g_{\mu\nu}\Box-\nabla_{\mu}\nabla_{\nu}\Big)\ln\left(\frac{\Box}{\mu^2}\right)R
    -\bar{\beta}\left({\delta^{\rho}}_{\mu}R_{\nu\sigma}+{\delta^{\rho}}_{\nu}R_{\mu\sigma}\right.\nonumber\\
    &-&\left.\frac{1}{2}g_{\mu\nu}{R^{\rho}}_{\sigma}+{\delta^{\rho}}_{\mu}g_{\nu\sigma}\Box+
   g_{\mu\nu}\nabla^{\rho}\nabla_{\sigma}-{\delta^{\rho}}_{\mu}\nabla_{\sigma}\nabla_{\nu}-{\delta^{\rho}}_{\nu}\nabla_{\sigma}\nabla_{\mu}\right)\ln\left(\frac{\Box}{\mu^2}\right){R^{\sigma}}_{\rho}.
\end{eqnarray}
We now solve the quantum corrected Einstein and Maxwell equations using perturbation theory in the gravitational coupling $G_N$. Classically, a four-dimensional charged AdS black hole is described by its mass $M$, its charge $Q$ and the AdS radius $L$, which is related to the cosmological constant via
\begin{equation}
    L=\sqrt{\frac{3}{\vert \Lambda \rvert}}.
\end{equation}
In turn, the cosmological constant is proportional to the vacuum energy density $\rho$ as  $\Lambda=8\pi G_N \rho$. For our future calculations we find convenient to keep the classical metric written in terms of $\rho$.
We now consider a small perturbation of order $\mathcal{O}\left(G^2_N\right)$ around the classical solution:
\begin{equation}\label{eq:expansion}
    g_{\mu\nu}=g^{AdS}_{\mu\nu}+g^{q}_{\mu\nu},
\end{equation}
where
\begin{eqnarray}
    ds^2_{AdS}&=&g^{AdS}_{\mu\nu}dx^{\mu}dx^{\nu},\nonumber\\
    &=&-\left(1-\frac{2G_N M}{r}+\frac{G_N Q^2}{r^2}+\frac{1}{3}8\pi\rho G_N r^2\right)dt^2,\nonumber\\
    &+&\left(1-\frac{2G_N M}{r}+\frac{G_N Q^2}{r^2}+\frac{1}{3}8\pi\rho G_N r^2\right)^{-1}dr^2+
    r^2d\theta^2+r^2\sin^2\theta d\phi^2.
\end{eqnarray}
The quantum effects are encoded in $g^q_{\mu\nu}$. We set $g_{\theta\theta}=g_{\phi\phi}=0$ and introduce two functions $\Sigma(r)$ and $\Omega(r)$ such that
\begin{equation}\label{eq:quantum_metric}
    ds^2_{q}=g^{q}_{\mu\nu}dx^{\mu}dx^{\nu}=G^2_N\Delta(r)dt^2+G^2_N\Sigma(r)dr^2.
\end{equation}
The full metric is then
\begin{equation}\label{eq:full_metric}
\begin{gathered}
    ds^2=g_{\mu\nu}dx^{\mu}dx^{\nu}\\=\bigg[-\Big(1-\frac{2G_N M}{r}+\frac{G_N Q^2}{r^2}+\frac{1}{3}8\pi\rho G_N r^2\Big)+G^2_N\Delta(r)\bigg]dt^2
   \\ +\bigg[\left(1-\frac{2G_N M}{r}+\frac{G_N Q^2}{r^2}+\frac{1}{3}8\pi\rho G_N r^2\right)^{-1}+G^2_N\Sigma(r)\bigg]dr^2+
    r^2d\theta^2+r^2\sin^2\theta d\phi^2.
\end{gathered}
\end{equation}
Quantum effects can also shift the classical value of the non-vanishing components of the electromagnetic tensor $F_{\mu\nu}$. Thus, we introduce a function $\Omega(r)$ such that
\begin{equation}\label{eq:expanded_F}
    F_{tr}=-F_{rt}=\frac{Q}{r^2}+G^2_N\Omega(r).
\end{equation}
Next, we plug the metric \eqref{eq:full_metric} and the components of the electromagnetic tensor \eqref{eq:expanded_F} into the equations \eqref{eq:maxwell_equations}, \eqref{eq:einstein_equations}, keeping only terms up to order $\mathcal{O}\left(G^2_N\right)$.
Since $H_{\mu\nu}$ and $K_{\mu\nu}$ are quadratic in curvature, they just need to be evaluated on the classical charged AdS solution, i.e. \eqref{eq:einstein_equations} can be expressed schematically as
\begin{equation}\label{eq:einstein_quantum}
\frac{1}{8\pi G_N}\left(G_{\mu\nu}-\Lambda g_{\mu\nu}\right)\left[g^{AdS}+g^{q}\right]+2 \left(H_{\mu\nu}\left[g^{AdS}\right]+K_{\mu\nu}\left[g^{AdS}\right]\right)=T_{\mu\nu}\left[g^{AdS}+g^{q}\right].
\end{equation}
The results of the action of $\ln\left(\Box/\mu^2\right)$ on several radial functions were collected in \cite{Delgado:2022pcc}. However, unlike the pure Reissner-Nordstr\"om solution, the charged AdS spacetime has the non-vanishing Ricci curvature $R=-32G\pi\rho$. Hence, we need to additionally compute the action of $\ln\left(\Box/\mu^2\right)$ on a constant. Using the formula \cite{Calmet:2019eof}
\begin{eqnarray}
    \ln\left(\frac{\Box}{\mu^2}\right)f(r)&=&\frac{1}{r}\int_0^{+\infty}dr'\,\frac{r'}{r+r'}f(r')-\lim_{\epsilon\to 0^{+}}\bigg\{ \frac{1}{r}\int_0^{r-\epsilon}dr'\,\frac{r'}{r-r'}f(r')\nonumber\\
    &+&\frac{1}{r}\int_{r+\epsilon}^{+\infty}dr'\,\frac{r'}{r'-r}f(r')
    +2f(r)\left[\gamma_E+\ln\left(\mu\epsilon\right)\right]\bigg\},\label{eq:formula_actionlog}
\end{eqnarray}
the result is
\begin{equation}
    \ln\left(\frac{\Box}{\mu^2}\right)\cdot 1=-2\left(\ln(\mu r) + \gamma_E -1 \right),
\end{equation}
where $\gamma_E$ is the Euler-Mascheroni constant.
The corresponding calculation is collected in Appendix \ref{sec:appendix1}, where we show how to regularise the integrals.
The final solution of the quantum corrected equations of motion \eqref{eq:einstein_quantum} and \eqref{eq:maxwell_equations} is
\begin{equation}\label{eq:final_metric}
\begin{gathered}
    ds^2=-f(r)dt^2+\frac{1}{g(r)}dr^2+r^2d\theta^2+r^2\sin^2\theta d\phi^2,
\end{gathered}
\end{equation}
with
\begin{eqnarray}
    f(r)&=&1-\frac{2G_N M}{r}+\frac{G_N Q^2}{r^2}+\frac{8\pi}{3}\rho G_N r^2+256\pi^2\rho G_N^2\left[4c_1+c_2+2(4\alpha+\beta)\ln(\mu r)\right]\nonumber\\
    &-&\frac{32\pi G^2_N Q^2}{r^4} \Big[c_2+4c_3+\left(\beta+4\gamma\right)\left(2\ln\left(\mu r\right)+2\gamma_E-3\right)\Big],\\
    g(r)&=&1-\frac{2G_N M}{r}+\frac{G_N Q^2}{r^2}+\frac{1}{3}8\pi\rho G_N r^2+512\pi^2\rho G_N^2(4\alpha+\beta)\nonumber\\
   & -&\frac{64\pi G^2_N Q^2}{r^4}\Big[c_2+4c_3+2\left(\beta+4\gamma\right)\left(\ln\left(\mu r\right)+\gamma_E-2\right)\Big], 
   \end{eqnarray}
and
\begin{eqnarray}
    F_{tr}=-F_{rt}&=&\frac{Q}{r^2}+\frac{16\pi G^2_N Q^3}{r^6}\Big[c_2+4c_3+\left(\beta+4\gamma\right)\left(2\ln\left(\mu r\right)+2\gamma_E-5\right)\Big]\nonumber\\
   & +&\frac{128\pi^2\rho G^2_NQ}{r^2}\left[4c_1+c_2+(4\alpha+\beta)\ln(\mu r)\right].
\end{eqnarray}
In the limit $\rho\to 0$ (or equivalently $L\to\infty$) one correctly recovers the solution obtained in \cite{Delgado:2022pcc} for the pure Reissner-Nordstr\"om black hole.
Although the metric seems to depend on the arbitrary energy scale $\mu$, the renormalised constants $c_1$, $c_2$, and $c_3$ also carry an explicit scale dependence \cite{El-Menoufi:2015cqw}:
\begin{equation}\label{eq:coefficients_scale}
\begin{split}
    c_1(\mu)=c_1(\mu_*)-\alpha\ln\left(\frac{\mu^2}{\mu^2_*}\right),\\
    c_2(\mu)=c_2(\mu_*)-\beta\ln\left(\frac{\mu^2}{\mu^2_*}\right),\\
    c_3(\mu)=c_3(\mu_*)-\gamma\ln\left(\frac{\mu^2}{\mu^2_*}\right),\\
\end{split}
\end{equation}
where $\mu_*$ is some fixed scale where the effective theory is matched onto the full theory. Taking into account \eqref{eq:coefficients_scale}, one sees that the metric ultimately does not depend on $\mu$, as it must be.

 %%%%%%%%%%%%%%%%%%%%%%%%%%
%%%%%%%%%%%%%%%%%%%%%%%%%%%%%%%%%%%%%%%%%%%%%%%%%
\section{Corrected entropy}\label{sec:entropy}
In this section we compute the quantum gravitational corrections to the entropy.  Analogously to what done in \cite{Delgado:2022pcc}, we consider a small electric charge. In addition, we assume the cosmological constant to be small, or, equivalently, the AdS radius to be large. We then compute the entropy up to orders $\mathcal{O}\left(Q^2\right),\mathcal{O}\left(\rho\right)$.
The corrections to the metric imply a shift to the classical outer horizon radius as
\begin{eqnarray}
    r_h&=&2G_NM-\frac{Q^2}{2M}+\frac{8\pi Q^2}{G_NM^3}\Big[c_2+4c_3+2\left(\beta+4\gamma\right)\left(\ln\left(2G_N M\mu \right)+\gamma_E-2\right)\Big]\nonumber\\
   & +&\frac{8192\pi^3G_NQ^2\rho}{M^3}(4\alpha+\beta)\Big[2c_2+8c_3+(\beta+4\gamma)\left(4\ln(2GM\mu)+4\gamma_E-9\right)\Big]\nonumber\\
   & +&\frac{512\pi^2G^2_NQ^2\rho}{3M}\Big[c_2+4c_3+(\beta+4\gamma)\left(2\ln(2G_N\mu)+2\gamma_E-5\right)\Big].\label{eq:new_radius}
\end{eqnarray}
We introduce the totally antisymmetric symbol $\epsilon_{\mu\nu}$
\begin{equation}
\epsilon_{\mu\nu}=
\begin{cases}
\sqrt{f(r)/g(r)} \hspace{10mm} \text{if} \hspace{1 mm} (\mu,\nu)=(t,r) \\
-\sqrt{f(r)/g(r)} \hspace{7.5 mm} \text{if} \hspace{1 mm} (\mu,\nu)=(r,t)\\
0 \hspace{30 mm} \text{otherwise},
\end{cases}
\end{equation}
so that the Wald formula for the entropy is \cite{Wald:1993nt}
\begin{eqnarray}
    S_{\text{Wald}}&=&-2\pi \int\limits_{r=r_h} d\Sigma\, \epsilon_{\mu\nu}\epsilon_{\rho\sigma}\frac{\partial \mathcal{L}}{\partial R_{\mu\nu\rho\sigma}},\nonumber\\
    &=& -8\pi\sqrt{\frac{f(r_h)}{g(r_h)}}\int\limits_{r=r_h}d\Sigma\, \frac{\partial \mathcal{L}}{\partial R_{rtrt}}
    =-8\pi\sqrt{\frac{f(r_h)}{g(r_h)}}\frac{\partial \mathcal{L}}{\partial R_{rtrt}}\bigg\rvert_{r=r_h}4\pi r^2_h.\label{eq:wald}
\end{eqnarray}
All the terms in the Lagrangian have to be considered without invoking the Gauss-Bonnet theorem. Furthermore, the following relations are useful:
\begin{eqnarray}
    \frac{\partial}{\partial R_{\mu\nu\rho\sigma}} R&=&\frac{1}{2}\left(g^{\mu\rho}g^{\nu\sigma}-g^{\mu\sigma}g^{\nu\rho}\right),\\
     \frac{\partial}{\partial R_{\mu\nu\rho\sigma}}R_{\alpha\beta}R^{\alpha\beta}&=&\frac{1}{2}\bigg(g^{\mu\rho}R^{\nu\sigma}-g^{\nu\rho}R^{\mu\sigma}-g^{\mu\sigma}R^{\nu\rho}+g^{\nu\sigma}R^{\mu\rho}\bigg).\\
    \frac{\partial}{\partial R_{\mu\nu\rho\sigma}}
   R_{\alpha\beta\gamma\delta}R^{\alpha\beta\gamma\delta}&=&2R^{\mu\nu\rho\sigma}.
\end{eqnarray}
In applying the formula one also needs the results of the action of  $\ln\left(\Box/\mu^2\right)$ on some radial functions. These were computed in \cite{Delgado:2022pcc}. However, the Riemann and Ricci tensors obtained from \eqref{eq:final_metric} contain new terms proportional to $\rho$.  In particular, one needs to know the action of  $\ln\left(\Box/\mu^2\right)$ on $1/r$ and $1/r^2$. The corresponding calculations are collected in Appendix \ref{sec:appendix1}.
The final result for the entropy of a charged AdS black hole is, up to order $\mathcal{O}\left(Q^4,\rho^2\right)$,
\begin{equation}\label{eq:corrected_entropy}
    S_{\text{Wald}}=\frac{A}{4G}-2\pi Q^2+S_{Sch}+S_{RN}+S_{AdS}+\mathcal{O}\left(Q^4,\rho^2\right),
\end{equation}
where $A=16\pi G^2M^2$ is the classical area of the event horizon of a Schwarzschild black
hole, $S_{Sch}$ is the quantum correction to the entropy of a Schwarzschild black hole, $S_{RN}$ represents the corrections to the entropy of a Reissner-Nordstr\"om black hole and $S_{AdS}$ are the new corrections containing the vacuum energy density $\rho$. The expression of $S_{Sch}$ is \cite{Delgado:2022pcc, Calmet:2021lny}
\begin{equation}
    S_{Sch}=64\pi^2c_3+64\pi^2\gamma\Big[4\ln\left(2G_NM\mu\right)+2\gamma_E-2\Big].
\end{equation}
The expression of $S_{RN}$ is \cite{Delgado:2022pcc}
\begin{eqnarray}
    S_{RN}&=&\frac{4\pi^2 Q^2}{G_NM^2}\left[ 5(c_2+4c_3)+\beta\left(10\gamma_E-21\right)+8\gamma\left(5\gamma_E-11\right)+10\left(\beta+4\gamma\right)\ln\left(2G_NM\mu\right)\right]\nonumber\\
   &+&\frac{64\pi^3Q^2}{9G^2_NM^4}\left\{54(\beta+4\gamma)\left[c_1+2\alpha\ln\left(2G_NM\mu\right)\right]
   +12(12\gamma_E-23)\beta\left[c_2+2\beta\ln\left(2G_NM\mu\right)\right]\right.\nonumber\\
   &+&\left. 48(48\gamma_E-97)\gamma\left[c_3+2\gamma\ln\left(2G_NM\mu\right)\right]
   +6\left[c_2\gamma(96\gamma_E-185)+c_3\beta(96\gamma_E-193)\right.\right.\nonumber\\
   &+&\left. \left. 12\beta\gamma(32\gamma_E-63)\ln\left(2G_NM\mu\right)\right]
   +36\left[c^2_2+4c_2\beta\ln\left(2G_NM\mu\right)+4\beta^2\ln^2\left(2G_NM\mu\right)\right]\right.\nonumber\\
   &+&\left. 576\left[c^2_3+4c_3\gamma\ln\left(2G_NM\mu\right)+4\gamma^2\ln^2\left(2G_NM\mu\right)\right]
   +288\left[c_2c_3+2c_2\gamma\ln\left(2G_NM\mu\right)\right.\right.\nonumber\\
   &+&\left. \left. 2c_3\beta\ln\left(2G_NM\mu\right)
   +4\beta\gamma\ln^2\left(2G_NM\mu\right)\right]+(\beta+4\gamma)\left[ 9\alpha(12\gamma_E-25)
  \right. \right.\nonumber\\
  &+&\left. \left. 8\beta\left(3\gamma_E(6\gamma_E-23)+40+3\pi^2\right)
   +4\gamma\big(6\gamma_E\left(24\gamma_E-97\right)+331+30\pi^2\big)\right]
\right\}.\label{eq:S_2}
\end{eqnarray}
The expression of $S_{AdS}$ is quite lengthy and it is relegated in Appendix \ref{sec:appendix2}.
Inserting the explicit scale dependence of the coefficients according to \eqref{eq:coefficients_scale}, one can check with any computer software that
\begin{equation}
    \frac{\partial S_{\text{Wald}}}{\partial \mu}=0,
\end{equation}
i.e. the entropy is RG invariant. The $\mu$-independent entropy is then effectively obtained with the substitutions $\mu\to\mu_*$ and $c_i\to c_i(\mu_*)$ everywhere.

%%%%%%%%%%%%%%%%%%%%%%%%%%%%%%%%%%%%%%%%%%%%%%%%%%
%%%%%%%%%%%%%%%%%%%%%%%%%%%%%%%%%%%%%%%%%%%%%%%%%%
\section{Thermodynamics}\label{sec:thermodynamics}
In this section we compute the quantum gravitational corrections to relevant thermodynamics quantities, namely temperature, pressure, specific heat and Helmholtz free energy. The first law of thermodynamics for a black hole with mass $M$ and charge $Q$ is given by
\begin{equation}
    dM=TdS-Qd\Phi+PdV,
\end{equation}
where $P$, $T$, $V$, $\Phi$ are the pressure, the temperature, the volume and the electric potential, respectively.
The electric potential is 
\begin{eqnarray}
\Phi&=&\int_{r_h}^{+\infty} dr'\, F_{tr}=\int_{r_h}^{+\infty} dr'\, \Big(\frac{Q}{{r'}^2}+G^2_N\Omega(r')\Big),\nonumber\\
&=&\frac{Q}{2G_N M}+
\frac{64\pi^2\rho GQ}{M}\left[4c_1+c_2+ (8\alpha+2\beta)\left(1+\ln(2G_NM\mu)\right)\right]+\mathcal{O}\left(Q^3,\rho^2\right).
\end{eqnarray}
The temperature is given by
\begin{eqnarray}
    T&=&\frac{1}{4\pi}\sqrt{\frac{df(r)}{dr}\frac{dg(r)}{dr}}\bigg\rvert_{r=r_h},\nonumber
    \\&=&\frac{1}{8\pi G_N M}+\frac{8}{3}\rho M G_N^2-\frac{2\rho G_NQ^2}{3M}+T_{RN}+T_{AdS}+\mathcal{O}\left(Q^4,\rho^2\right),
\end{eqnarray}
where $T_{RN}$ represents the quantum gravitational corrections to a Reissner-Nordstr\"om black hole and $T_{AdS}$ contains the new corrections proportional to $\rho$. The corresponding expressions are
\begin{equation}
\begin{gathered}
   T_{RN}=\frac{Q^2}{4G^3_NM^5}\Big[2(c_2+4c_3)+(\beta+4\gamma)\left(4\gamma_E-9+4\ln(2G_NM\mu)\right)\Big],
\end{gathered}
\end{equation}
and
\begin{eqnarray}
    T_{AdS}&=&\frac{32\pi \rho G_N}{M}(4\alpha+\beta) -\frac{8\pi \rho Q^2}{3M^3}\Big[4(c_2+4c_3)-12\alpha-(27-8\gamma_E)\beta+32(-3+\gamma_E)\gamma
    \nonumber\\
    &+&8(\beta+4\gamma)\ln(2G_NM\mu)\Big] -\frac{128\pi^2\rho Q^2}{G_NM^5}(4\alpha+\beta)\Big[17(c_2+4c_3) \nonumber\\
    &+&2(\beta+4\gamma)\left(17\gamma_E-38+17\ln(2G_NM\mu)\right)\Big].
\end{eqnarray}
This relation utilizes the behavior of the temperature of a AdS Reissner-Nordstr\"om black hole as a function of $M$.  
Quantum effects also produce corrections to the pressure \cite{Calmet:2021lny, Delgado:2022pcc}. For a charged AdS$_4$ black hole, the pressure is
\begin{eqnarray}
    P&=&-\frac{T\frac{dS}{dM}-Q\frac{d\Phi}{dM}-1}{\frac{dV}{dM}},\nonumber\\
    &=&-\frac{Q^2}{64\pi G^4_N M^4}-\frac{2}{3}\rho+P_{sch}+P_{RN}+P_{AdS}+\mathcal{O}\left(Q^4,\rho^2\right),
\end{eqnarray}
where $V=4/3\pi r^3_h$ is the volume, $P_{sch}=-\gamma/(G^4_NM^4)$ is the quantum correction for a Schwarzschild black hole, $P_{RN}$ corresponds to the corrections for a Reissner-Nordstr\"om black hole and $P_{AdS}$ represents the new contributions containing the AdS$_4$ radius.
The expression of $P_{RN}$ is \cite{Delgado:2022pcc}
\begin{multline}
    P_{RN}=\frac{Q^2}{32G^5_NM^6}\Big[c_2+4c_3+2\beta(\gamma_E-4)+8\gamma(\gamma_E-5)+2(\beta+4\gamma)\ln(2G_NM\mu)\Big]
    \\+\frac{\pi Q^2}{9G^6_NM^8}\bigg\{54(\beta+4\gamma)\Big[c_1+2\alpha\ln\left(2G_NM\mu\right)\Big]
    +24(6\gamma_E-13)\beta\Big[c_2+2\beta\ln\left(2G_NM\mu\right)\Big]\\
    +768(3\gamma_E-7)\gamma\Big[c_3+2\gamma\ln\left(2G_NM\mu\right)\Big]
    +6\Big[c_2\gamma(96\gamma_E-215)+c_3\beta(96\gamma_E-217)\\
    +96\beta\gamma(4\gamma_E-9)\ln\left(2G_NM\mu\right)\Big]
     +36\Big[c^2_2+4c_2\beta\ln\left(2G_NM\mu\right)+4\beta^2\ln^2\left(2G_NM\mu\right)\Big]\\
   +576\Big[c^2_3+4c_3\gamma\ln\left(2G_NM\mu\right)+4\gamma^2\ln^2\left(2G_NM\mu\right)\Big]
   +288\Big[c_2c_3+2c_2\gamma\ln\left(2G_NM\mu\right)\\
   +2c_3\beta\ln\left(2G_NM\mu\right)
    +4\beta\gamma\ln^2\left(2G_NM\mu\right)\Big]+(\beta+4\gamma)\Big[36\alpha(3\gamma_E-7)\\ +2\beta\big(8\gamma_E(9\gamma_E-39)+229+12\pi^2\big)
    +\gamma\big(192\gamma_E(3\gamma_E-14)+2095+120\pi^2\big)\Big]  
    \bigg\}.
    \end{multline}
The expression of $P_{AdS}$ is reported in Appendix \ref{sec:appendix2}.
Plugging the explicit scale dependence \eqref{eq:coefficients_scale}, it is easy to verify with any computer software that the pressure is RG invariant,
\begin{equation}
    \frac{\partial P}{\partial \mu}=0.
\end{equation}
Next, we compute the specific heat which is an important quantity related to the thermodynamics stability. The black hole is stable if $C\geq0$ (see Fig.4). One obtains
\begin{eqnarray}
    C&=&\frac{T}{M}\frac{dS}{dT}=\frac{T}{M}\frac{1}{\frac{dT}{dM}}\frac{dS}{dM}=-8\pi GM+C_{Sch}+C_{RN}+C_{AdS}+\mathcal{O}\left(Q^4,\rho^2\right),
\end{eqnarray}
where $C_{Sch}$ is the quantum contribution to a Schwarzschild black hole given as
\begin{equation}
   C_{Sch}=-\frac{128\pi^2\gamma}{M},
\end{equation}
and $ C_{RN}$ is the quantum correction to a Reissner-Nordstr\"om black hole   given as
\begin{eqnarray}
   C_{RN}&=&\frac{8\pi^2Q^2}{G_NM^3}\left[ 21(c_2+4c_3)+2\beta(21\gamma_E-53)+4\gamma(42\gamma_E-107)+42(\beta+4\gamma)\ln(2GM\mu)\right]\nonumber\\
   &+&\frac{512\pi^3 Q^2}{9G^2M^5}\bigg\{18 c_2^2+288 c_3^2+3 c_3 \left[(-217+96 \gamma_E) \beta +8 (-103+48 \gamma_E) \gamma \right]\nonumber\\
   &+&3 c_2 \left[48 c_3+4 (-13+6 \gamma_E) \beta +(-197+96 \gamma_E) \gamma \right]+(\beta +4 \gamma )\left[27 c_1+229 \beta +773 \gamma\right.\nonumber\\
   &+&6\left.\left[3 (-7+3 \gamma_E) \alpha +12 \gamma_E^2 (\beta +4 \gamma )+2 \pi ^2 (\beta +5 \gamma )-2 \gamma_E (26 \beta +103 \gamma )\right]\right]\nonumber\\
   &+&6 (\beta +4 \gamma ) \ln\left(2G_N M\mu\right) \left[12 c_2+48 c_3+9 \alpha +(-52 +24 \gamma_E) \beta\right.\nonumber\\
   &+&\left( -206 +96 \gamma_E) \gamma +12 (\beta +4 \gamma ) \ln\left(2G_N M\mu\right)\right]\bigg\}.
\end{eqnarray}
 The purely AdS part is given in Appendix \ref{sec:appendix2}.
 
 Now let us calculate the Helmholtz free energy  as
\begin{eqnarray}
    F&=&E-TS=M+Q\Phi-TS\nonumber\\
    &=&\frac{M}{2}+\frac{3Q^2}{4G_NM}+8\pi G^2_NMQ^2\rho+F_{Sch}+F_{RN}+F_{AdS}+\mathcal{O}\left(Q^3,\rho^2\right),
\end{eqnarray}
with
\begin{equation}
    F_{Sch}=-\frac{8\pi}{G_NM}\Big[c_3+2\gamma\left(\ln(2G_NM\mu)+\gamma_E-1\right)\Big],
\end{equation}
and
\begin{equation}
\begin{gathered}
F_{RN}=-\frac{\pi Q^2}{2G^2_NM^3}\Big[9(c_2+4c_3)+3\beta(6\gamma_E-13)+8\gamma(9\gamma_E-20)+18(\beta+4\gamma)\ln(2G_NM\mu)\Big]\\
-\frac{8\pi^2Q^2}{9G^3_NM^5}\bigg\{36 c_2^2+720 c_3^2+6 c_2 [54 c_3+(-46+24 \gamma_E) \beta +(-197+108 \gamma_E) \gamma ]\\
+24 c_3 \Big[(-55+27 \gamma_E) \beta +(-233+120 \gamma_E) \gamma \Big]+(\beta +4 \gamma )\Big[54 c_1+9 (-25+12 \gamma_E) \alpha\\
 +8 \left[40+3 \gamma_E (-23+6 \gamma_E)+3 \pi ^2\right] \beta 
+4 \left[412+3 \gamma_E (-233+60 \gamma_E)+30 \pi ^2\right] \gamma \Big]\\
+12 \ln\left(2G_N M\mu\right)\Big[(\beta +4 \gamma ) \left[9 \alpha +(-46+24 \gamma_E) \beta +(-233+120 \gamma_E) \gamma \right]\\
+6 c_2 (2 \beta +9 \gamma )+6 c_3 (9 \beta +40 \gamma )
+12 (\beta +4 \gamma ) (\beta +5 \gamma ) \ln\left(2G_N M\mu\right)\Big] \bigg\}.
\end{gathered}
\end{equation}
The expression for $F_{AdS}$ is reported in Appendix \ref{sec:appendix2}.
Again, one can verify that both specific heat and Helmholtz free energy are RG invariant as
\begin{equation}
    \frac{\partial C}{\partial \mu}=0, \hspace{5mm} \frac{\partial F}{\partial \mu}=0.
\end{equation}
%%%%%%%%%%%%%%%%%%%%%%%%%%%%%%%%%%%%%%%%%%%%%%%%%%
%%%%%%%%%%%%%%%%%%%%%%%%%%%%%%%%%%%%%%%%%%%%%%%%%%
\section{Conclusions and outlook}\label{sec:conclusions}
In this study, we have employed the effective field theory framework for quantum gravity to rigorously compute quantum gravitational corrections to the entropy of charged AdS black holes. Our calculations have specifically accounted for contributions up to the second order in curvature, ensuring a systematic and comprehensive approach to capturing higher-order effects. By explicitly incorporating the scale dependence of the coefficients in our analysis, we have established that the Wald entropy, a key thermodynamic quantity, exhibits RG invariance. This invariance underscores the robustness of the entropy calculations under changes in the energy scale, providing a critical consistency check for the theoretical framework.

In addition to entropy, we have derived quantum gravitational corrections to several fundamental thermodynamic quantities associated with black holes, including temperature, pressure, specific heat, and Helmholtz free energy. These quantities have been shown to satisfy the first law of thermodynamics, thereby reinforcing the internal consistency of the corrections within the thermodynamic framework. Furthermore, our detailed examination has revealed that specific heat and Helmholtz free energy, in particular, maintain their RG invariance even after the inclusion of quantum gravitational effects. This finding is significant, as it confirms the stability of these thermodynamic properties under scale transformations and highlights their compatibility with the principles of effective field theory.

Overall, our results have provided new insights into the interplay between quantum gravity and black hole thermodynamics. By demonstrating the RG invariance of critical quantities and ensuring adherence to established thermodynamic laws, we have validated the applicability and reliability of the effective field theory approach in this context. These findings contribute to a deeper understanding of charged AdS black hole physics and offer a solid foundation for further explorations into the quantum aspects of gravity.
 


\bibliographystyle{ieeetr}
\bibliography{references}

\begin{thebibliography}{10}

\bibitem{PhysRevD.7.2333}
J.~D. Bekenstein, ``Black holes and entropy,'' {\em Phys. Rev. D}, vol.~7,
  pp.~2333--2346, Apr 1973.

\bibitem{PhysRevD.104.066012}
X.~Calmet and F.~Kuipers, ``Quantum gravitational corrections to the entropy of
  a schwarzschild black hole,'' {\em Phys. Rev. D}, vol.~104, p.~066012, Sep
  2021.

\bibitem{PhysRevD.48.R3427}
R.~M. Wald, ``Black hole entropy is the noether charge,'' {\em Phys. Rev. D},
  vol.~48, pp.~R3427--R3431, Oct 1993.

\bibitem{PAUL2023116259}
P.~Paul, S.~Upadhyay, Y.~Myrzakulov, D.~V. Singh, and K.~Myrzakulov, ``More
  exact thermodynamics of nonlinear charged ads black holes in 4d critical
  gravity,'' {\em Nuclear Physics B}, vol.~993, p.~116259, 2023.

\bibitem{Kumar_2023}
J.~Kumar, S.~Upadhyay, and H.~K. Sudhanshu, ``Small black string
  thermodynamics,'' {\em Physica Scripta}, vol.~98, no.~9, p.~095306, 2023.

\bibitem{PhysRevLett.26.1344}
S.~W. Hawking, ``Gravitational radiation from colliding black holes,'' {\em
  Phys. Rev. Lett.}, vol.~26, pp.~1344--1346, May 1971.

\bibitem{PhysRevD.9.3292}
J.~D. Bekenstein, ``Generalized second law of thermodynamics in black-hole
  physics,'' {\em Phys. Rev. D}, vol.~9, pp.~3292--3300, Jun 1974.

\bibitem{Buchbinder:1992rb}
I.~L. Buchbinder, S.~D. Odintsov, and I.~L. Shapiro, {\em {Effective action in
  quantum gravity}}.
\newblock {Taylor \& Francis}, 1992.

\bibitem{Donoghue:2017pgk}
J.~F. Donoghue, M.~M. Ivanov, and A.~Shkerin, ``{EPFL Lectures on General
  Relativity as a Quantum Field Theory},'' 2 2017.

\bibitem{Myrzakulov:2014hca}
R.~Myrzakulov, S.~Odintsov, and L.~Sebastiani, ``{Inflationary universe from
  higher-derivative quantum gravity},'' {\em Phys. Rev. D}, vol.~91, no.~8,
  p.~083529, 2015.

\bibitem{Elizalde:2017mrn}
E.~Elizalde, S.~D. Odintsov, L.~Sebastiani, and R.~Myrzakulov,
  ``{Beyond-one-loop quantum gravity action yielding both inflation and
  late-time acceleration},'' {\em Nucl. Phys. B}, vol.~921, pp.~411--435, 2017.

\bibitem{Weinberg:1980gg}
S.~Weinberg, ``{Ultraviolet divergences in quantum theories of gravitation},''
  in {\em {General Relativity}: {An Einstein Centenary Survey}}, (Cambridge,
  UK), pp.~790--831, Univ. Pr., 1980.

\bibitem{Starobinky:1981ZhPmR}
A.~A. Starobinsky, ``{Evolution of small perturbations of isotropic
  cosmological models with one-loop quantum gravitational corrections},'' {\em
  JETP Lett.}, vol.~34, pp.~438--441, 1981.

\bibitem{Barvinsky:1983vpp}
A.~O. Barvinsky and G.~A. Vilkovisky, ``{The generalized Schwinger-DeWitt
  technique and the unique effective action in quantum gravity},'' {\em Phys.
  Lett. B}, vol.~131, pp.~313--318, 1983.

\bibitem{Barvinsky:1985an}
A.~O. Barvinsky and G.~A. Vilkovisky, ``{The Generalized Schwinger-DeWitt
  Technique in Gauge Theories and Quantum Gravity},'' {\em Phys. Rept.},
  vol.~119, pp.~1--74, 1985.

\bibitem{Barvinsky:1987uw}
A.~O. Barvinsky and G.~A. Vilkovisky, ``{Beyond the Schwinger-Dewitt Technique:
  Converting Loops Into Trees and In-In Currents},'' {\em Nucl. Phys. B},
  vol.~282, pp.~163--188, 1987.

\bibitem{Barvinsky:1990up}
A.~O. Barvinsky and G.~A. Vilkovisky, ``{Covariant perturbation theory. 2:
  Second order in the curvature. General algorithms},'' {\em Nucl. Phys. B},
  vol.~333, pp.~471--511, 1990.

\bibitem{Donoghue:1994dn}
J.~F. Donoghue, ``{General relativity as an effective field theory: The leading
  quantum corrections},'' {\em Phys. Rev. D}, vol.~50, pp.~3874--3888, 1994.

\bibitem{Donoghue:2014yha}
J.~F. Donoghue and B.~K. El-Menoufi, ``{Nonlocal quantum effects in cosmology:
  Quantum memory, nonlocal FLRW equations, and singularity avoidance},'' {\em
  Phys. Rev. D}, vol.~89, no.~10, p.~104062, 2014.

\bibitem{Donoghue:2015nba}
J.~F. Donoghue and B.~K. El-Menoufi, ``{Covariant non-local action for massless
  QED and the curvature expansion},'' {\em JHEP}, vol.~10, p.~044, 2015.

\bibitem{Calmet:2018elv}
X.~Calmet, ``{Vanishing of Quantum Gravitational Corrections to Vacuum
  Solutions of General Relativity at Second Order in Curvature},'' {\em Phys.
  Lett. B}, vol.~787, pp.~36--38, 2018.

\bibitem{Delgado:2022pcc}
R.~C. Delgado, ``{Quantum gravitational corrections to the entropy of a
  Reissner\textendash{}Nordstr\"om black hole},'' {\em Eur. Phys. J. C},
  vol.~82, no.~3, p.~272, 2022.
\newblock [Erratum: Eur.Phys.J.C 83, 468 (2023)].

\bibitem{Calmet:2019eof}
X.~Calmet, R.~Casadio, and F.~Kuipers, ``{Quantum Gravitational Corrections to
  a Star Metric and the Black Hole Limit},'' {\em Phys. Rev. D}, vol.~100,
  no.~8, p.~086010, 2019.

\bibitem{El-Menoufi:2015cqw}
B.~K. El-Menoufi, ``{Quantum gravity of Kerr-Schild spacetimes and the
  logarithmic correction to Schwarzschild black hole entropy},'' {\em JHEP},
  vol.~05, p.~035, 2016.

\bibitem{Wald:1993nt}
R.~M. Wald, ``{Black hole entropy is the Noether charge},'' {\em Phys. Rev. D},
  vol.~48, no.~8, pp.~R3427--R3431, 1993.

\bibitem{Calmet:2021lny}
X.~Calmet and F.~Kuipers, ``{Quantum gravitational corrections to the entropy
  of a Schwarzschild black hole},'' {\em Phys. Rev. D}, vol.~104, no.~6,
  p.~066012, 2021.

\end{thebibliography}

\appendix
\section{Action of $\ln\left(\frac{\Box}{\mu^2}\right)$ on radial functions}\label{sec:appendix1}
We show the result of the action of $\ln\left(\Box/\mu^2\right)$ on some radial functions. Let us start with the constant function. We regularize the integrals in \eqref{eq:formula_actionlog} as
\begin{equation}
\begin{gathered}
    \ln\left(\frac{\Box}{\mu^2}\right)1=\lim_{\epsilon\to 0}\bigg[\frac{1}{r}\int_0^{1/\sqrt{\epsilon/r^3}}dr'\,\frac{r'}{r+r'}\\- \frac{1}{r}\int_0^{r-\sqrt{\epsilon r}}dr'\,\frac{r'}{r-r'}
    -\frac{1}{r}\int_{r+\sqrt{\epsilon r}}^{+\infty}dr'\,\frac{r'}{r'-r} -2\gamma_E-2\ln\left(\mu\epsilon\right)\bigg],\\
\end{gathered}
\end{equation}
where we have added the $r$ in the integral limits for dimensional reasons (they must have the dimension of a length). The divergences cancel out:
\begin{equation}
\begin{gathered}
     \ln\left(\frac{\Box}{\mu^2}\right)1=\lim_{\epsilon\to 0}\Big[2+\ln(\epsilon r)\\-\ln\left(-r+\sqrt{r^3/\epsilon}\right)-\ln\left(r+\sqrt{r^3/\epsilon}\right)-2\gamma_E-2\ln(\mu\epsilon)\Big]\\=-2(\ln(\mu r)+\gamma_E-1).
\end{gathered}
\end{equation}
Let us consider now $f(r)=1/r$. We regularize the integrals in \eqref{eq:formula_actionlog} as
\begin{equation}
\begin{gathered}
    \ln\left(\frac{\Box}{\mu^2}\right)\frac{1}{r}=\lim_{M\to\infty}\lim_{\epsilon\to0}\bigg\{\frac{1}{r}\int_0^M dr^{'}\frac{r'}{r+r'}\frac{1}{r'}\\-\frac{1}{r}\int_0^{r-\epsilon}dr'\frac{r'}{r-r'}\frac{1}{r'} -\frac{1}{r}\int_{r+\epsilon}^M dr'\frac{r'}{r'-r}\frac{1}{r'}-\frac{2}{r}\left[\gamma_E+\ln\left(\mu\epsilon\right)\right]\bigg\}.
\end{gathered}
\end{equation}
Again, the divergences cancel:
\begin{equation}
\begin{gathered}
\ln\left(\frac{\Box}{\mu^2}\right)\frac{1}{r}=\lim_{M\to\infty}\lim_{\epsilon\to0}\frac{1}{r}\bigg[\ln\left(\frac{M}{r}\right)-\ln\left(\frac{M}{\epsilon}\right)\\-\ln\left(\frac{r}{\epsilon}\right)
-2\gamma_E-2\ln(\mu\epsilon)\bigg]=-\frac{2}{r}\left(\ln(\mu r)+\gamma_E\right).
\end{gathered}
\end{equation}
Analogously,
\begin{equation}
\begin{gathered}
    \ln\left(\frac{\Box}{\mu^2}\right)\frac{1}{r^2}=\text{Re}\lim_{\epsilon\to0}\bigg\{\frac{1}{r}\int_0^\infty dr^{'}\frac{r'}{r+r'}\frac{1}{{(r'+i\epsilon})^2}\\-\frac{1}{r}\int_0^{r-\epsilon}dr'\frac{r'}{r-r'}\frac{1}{{(r'+i\epsilon)}^2}
    -\frac{1}{r}\int_{r+\epsilon}^\infty dr'\frac{r'}{r'-r}\frac{1}{{r'}^2}-\frac{2}{r^2}\left[\gamma_E+\ln\left(\mu\epsilon\right)\right]\bigg\}\\=-\frac{2}{r^2}\left(\ln\left(\mu r\right)+\gamma_E\right).
\end{gathered}
\end{equation}
%%%%%%%%%%%%%%%%%%%%%%%%%%%%%%%%%%%%%%%%%%%%%%%%%%%%%%%%%%%%
%%%%%%%%%%%%%%%%%%%%%%%%%%%%%%%%%%%%%%%%%%%%%%%%%%%%%%%%%%%%
\section{Explicit expressions of the $\rho$-dependent quantum corrections}\label{sec:appendix2}
We provide the explicit expressions of the $\rho$ - dependent quantum corrections for several thermodynamics quantities discussed in the paper. We checked with Mathematica that all these quantities are RG invariant. 

The correction to the entropy is 
\begin{equation}
\begin{gathered}
S_{AdS}=-16384  \pi ^4\rho G_N^2 (4 \alpha +\beta )  \Big[3 c_1-c_3+(6\alpha-2\gamma) \left(\gamma_E+\ln\left(2 G_N M \mu\right)\right)\Big] \\
+\frac{256}{3} \pi^3 \rho  G_N^2 Q^2 \bigg\{36 c_1+17 c_2+36 c_3+12 (-3+4 \gamma_E) \alpha +7 (-7+4 \gamma_E) \beta \\
+4 (-41+18 \gamma_E) \gamma +(72 \alpha +34 \beta +72 \gamma ) \ln\left(2 G_N M \mu\right)\bigg\}\\
-\frac{131072 \pi ^5 \rho Q^2 }{M^4}(4 \alpha +\beta )  \bigg\{c_2 \Big[4 c_3+3 \alpha +(-17+8 \gamma_E) \gamma \Big]
+2 \Big[8 c_3^2\\+c_3 [6 \alpha +(-9 +4 \gamma_E) \beta +( -70 +32 \gamma_E) \gamma ]+(-2+\gamma_E) (\beta +4 \gamma ) [3 \alpha +(-19+8 \gamma_E) \gamma ]\Big]\\
+2 \ln\left(2 G_N M \mu\right) \Big[4 c_2 \gamma +4 c_3 (\beta +8 \gamma )+(\beta +4 \gamma ) (3 \alpha -35 \gamma +16 \gamma_E \gamma )\\+8 \gamma  (\beta +4 \gamma ) \ln\left(2 G_N M \mu\right)\Big]\bigg\}
-\frac{4096\pi ^4 \rho G_N Q^2}{3 M^2}  \bigg\{9 c_2^2+32 c_3^2\\-6 (4 \alpha +\beta ) \left[3 \alpha +(-42+(27-4 \gamma_E) \gamma_E) \beta \right]\\
+2 \Big[12 (169+4 \gamma_E (-27+4 \gamma_E)) \alpha +[555+4 \gamma_E (-95+16 \gamma_E)] \beta \Big] \gamma\\
+64 (-2+\gamma_E) (-3+2 \gamma_E) \gamma ^2+c_2 \Big[44 c_3+12 (-19+4 \gamma_E) \alpha +(-93+30\gamma_E)\beta\\+(-164+88\gamma_E)\gamma \Big]
+8 c_3 \Big[6 (-19+4 \gamma_E) \alpha +(-33+8 \gamma_E) \beta +4 (-7+4 \gamma_E) \gamma \Big]\\
+36 c_1 \Big[c_2+4 c_3+2 (-2+\gamma_E) (\beta +4 \gamma )\Big]\\
+2 \ln\left(2 G_N M \mu\right) \Big[4 c_3 (36 \alpha +11 \beta +16 \gamma )+2 c_2 (18 \alpha +9 \beta +22 \gamma )\\
+(\beta +4 \gamma ) \left[36 c_1+3 (-31+10 \gamma_E) (4 \alpha +\beta )+8 (-7+4 \gamma_E) \gamma \right]\\
+2 (\beta +4 \gamma ) (36 \alpha +9 \beta +8 \gamma ) \ln\left(2 G_N M \mu\right)\Big]\bigg\}.
\end{gathered}
\end{equation}
The correction to the pressure is
\begin{equation}
\begin{gathered}
P_{AdS}=\rho\bigg\{\frac{128 \pi ^2 (4 \alpha +\beta ) (3 \alpha -2 \gamma ) }{G_N^2 M^4}-\frac{8 \pi  (12 \alpha +3 \beta +4 \gamma ) }{3 G_N M^2}\bigg\}\\
+\frac{2 \pi \rho Q^2}{3 G_N^2 M^4}\bigg\{ -2(6 c_1+ c_2+10 c_3)-60 \alpha +(-49+10\gamma_E)\beta +8 (-18+5 \gamma_E) \gamma \\
+4 (-6 \alpha +\beta +10 \gamma ) \ln\left(2 G_N M \mu\right)\bigg\}+\frac{8 \pi^2 \rho  Q^2}{27 G_N^3 M^6}  \bigg\{-36 c_2^2\\
+3 c_2 \Big[240 c_3+(5652-576 \gamma_E) \alpha +(1121+24 \gamma_E) \beta  +160 (-5+3 \gamma_E) \gamma\Big]\\
 -432 c_1 \Big[3 c_2+12 c_3+2 (-8+3 \gamma_E) (\beta +4 \gamma )\Big]\\
 +2 \Big[1728 c_3^2+1944 \alpha ^2+\left[-8509+3 \gamma_E (941+48 \gamma_E)+96 \pi ^2\right] \beta ^2\\
 +12 \left[-2474+3 \gamma_E (99+64 \gamma_E)+72 \pi ^2\right] \beta  \gamma +64 \left[292+3 \gamma_E (-161+36 \gamma_E)+30 \pi ^2\right] \gamma ^2\\
 -18 \alpha  \left[(2327+6 \gamma_E (-201+16 \gamma_E)) \beta +8 \left(1186-603 \gamma_E+48 \gamma_E^2\right) \gamma \right]\\
 +6 c_3 \left[12 (471-48 \gamma_E) \alpha +677 \beta -2576 \gamma +192 \gamma_E (\beta +6 \gamma )\right]
\Big]\\
+6 \ln\left(2 G_N M \mu\right) \Big[-48 c_3 (36 \alpha -5 \beta -48 \gamma )-24 c_2 (18 \alpha +\beta -10 \gamma )\\
+(\beta +4 \gamma ) \Big[-432 c_1+36 (221-40 \gamma_E) \alpha +(1121+24 \gamma_E) \beta +16 (-161+72 \gamma_E) \gamma \Big]\\
-24 (36 \alpha +\beta -24 \gamma ) (\beta +4 \gamma ) \ln\left(2 G_N M \mu\right)\Big]\bigg\}+\frac{256 \pi ^3 \rho Q^2}{9 G_N^4 M^8}(4 \alpha +\beta )\bigg\{36 c_2^2\\
-576 c_3^2+6 c_2 \Big[-18 \alpha +4 (-13+6 \gamma_E) \beta +37 \gamma \Big]+6 c_3 \Big[-72 \alpha +23 \beta +16 (67-24 \gamma_E) \gamma \Big]\\
+(\beta +4 \gamma ) \Big[54 c_1+3 (51-36 \gamma_E) \alpha +458 \beta -4961 \gamma \\
+24 \left[6 \gamma_E^2 (\beta -4 \gamma )+\pi ^2 (\beta +5 \gamma )+\gamma_E (-26 \beta +134 \gamma )\right]\Big]\\
+12 \ln\left(2 G_N M \mu\right)\Big[12 c_2 \beta -192 c_3 \gamma +(\beta +4 \gamma ) \left[-9 \alpha -52 \beta +24 \gamma_E \beta +268 \gamma -96 \gamma_E \gamma \right]\\
+12 \left(\beta ^2-16 \gamma ^2\right) \ln\left(2 G_N M \mu\right)\Big]\bigg\}.
\end{gathered}
\end{equation}
The correction to the specific heat is 
\begin{equation}
\begin{gathered}
C_{AdS}=\frac{32768\pi^4 \rho G_N^2}{M}(4\alpha + \beta)(3\alpha-\gamma)\\
+\frac{512\pi ^3\rho G_N^2 Q^2}{3M}\bigg\{2\left(29c_2+116c_3-6\alpha\right)+(\beta +4 \gamma )\left[-283+116 \gamma_E+116\ln\left(2G_NM\mu\right)\right]\bigg\}\\
-\frac{262144\pi ^5\rho Q^2}{M^5} (4 \alpha +\beta )   \bigg\{32 c_3^2+4 c_2 [2 c_3+3 \alpha +(-1+4 \gamma_E) \gamma ]\\
+8 c_3 \Big[6 \alpha +(-5+2 \gamma_E) \beta +2 (-11+8 \gamma_E) \gamma \Big]\\
+(\beta +4 \gamma ) \Big[3 (-19+8 \gamma_E) \alpha +2 [9+4 \gamma_E (-11+4 \gamma_E)] \gamma \Big]\\
+8 \ln\left(2G_NM\mu\right) \Big[2 c_2 \gamma +2 c_3 (\beta +8 \gamma )+(\beta +4 \gamma ) (3 \alpha -11 \gamma +8 \gamma_E \gamma )\\+4 \gamma  (\beta +4 \gamma ) \ln\left(2G_NM\mu\right)\Big]\bigg\}
+\frac{8192\pi ^4\rho G_NQ^2}{27 M^3} \bigg\{
63c_2^2+72\Big[28c_3^2-12c_3(25+2\gamma_E)\alpha + 9\alpha^2\Big]\\
+6\Big[2c_3\left(-851+144\gamma_E\right)+4233\alpha-72\gamma_E(19+2\gamma_E)\alpha\Big]\beta\\
+\Big[8393+24\gamma_E(-194+15\gamma_E)+96\pi^2\Big]\beta^2
+24\Big[4 c_3 (-173+84 \gamma_E)\\+4206 \alpha -72 \gamma_E (19+2 \gamma_E) \alpha +\left(1577+6 \gamma_E (-187+24 \gamma_E)+36 \pi ^2\right) \beta \Big]\gamma\\
+16 \Big[1069+12 \gamma_E (-173+42 \gamma_E)+120 \pi ^2\Big] \gamma ^2\\
+3 c_2 \Big[252 c_3-72 (25+2 \gamma_E) \alpha +17 (-43+6 \gamma_E) \beta +8 (-113+63 \gamma_E) \gamma\Big]\\
-108 c_1 \Big[3 c_2+12 c_3+(-17+6 \gamma_E) (\beta +4 \gamma )\Big]+6 \ln\left(2G_NM\mu\right) \Big[12 c_3 (-36 \alpha +21 \beta +112 \gamma )\\
-(\beta +4 \gamma ) [108 c_1+36 (33+10 \gamma_E) \alpha +17 (43-6 \gamma_E) \beta +8 (173-84 \gamma_E) \gamma ]\\
+6 c_2 [-18 \alpha +7 (\beta +6 \gamma )]+6 (\beta +4 \gamma ) [-36 \alpha +7 (\beta +8 \gamma )] \ln\left(2G_NM\mu\right)\Big]\bigg\}.  
\end{gathered}
\end{equation}
The correction to the Helmholtz free energy is 
\begin{equation}
\begin{gathered}
F_{AdS}=\rho\bigg\{ \frac{2048\pi ^3 G_N}{M} (4 \alpha +\beta ) \Big[3 c_1-2 c_3+6 \gamma_E \alpha +(2 -4 \gamma_E) \gamma +(6 \alpha -4 \gamma ) \ln\left(2G_NM\mu\right)\Big]\\
  -\frac{128\pi ^2}{3} G_N^2 M  \Big[4 c_3+3 (4 \alpha +\beta )+8 (-1+\gamma_E) \gamma +8 \gamma  \ln\left(2G_NM\mu\right)\Big]
\bigg\}\\
 -\frac{32\pi ^2\rho G_N Q^2}{3M}\bigg\{12\left(c_1+c_2+3c_3\right)+48 (-2+\gamma_E) \alpha+(-61 +30 \gamma_E)\beta\\
+(-148 +72 \gamma_E) \gamma +24 (\alpha +\beta +3 \gamma ) \ln\left(2 G_N M \mu \right)\bigg\}\\
+\frac{128 \pi ^3 \rho Q^2}{27 M^3}\bigg\{36\Big[5c_2^2+c_2\left[16c_3+(-39+48\gamma_E)\alpha\right]-8\left[2c_3^2+3c_3(7-8\gamma_E)\alpha+\alpha^2\right]\Big]\bigg\}\\
+3\Big[32c_3(10+3\gamma_E)+c_2(-181+168\gamma_E)+24\left[83+3\gamma_E(-47+16\gamma_E)\right]\alpha\Big]\beta\\
+\Big[151+6\gamma_E(-37+48\gamma_E)-96\pi^2\Big]\beta^2
+24 \Big[c_3 (248-96 \gamma_E)+c_2 (-73+48 \gamma_E)\\+6 \left[199+96 (-3+\gamma_E) \gamma_E\right] \alpha +\left[43+6 \gamma_E (13+4 \gamma_E)-36 \pi ^2\right] \beta \Big] \gamma\\
 -64 \Big[7+6 \gamma_E (-31+6 \gamma_E)+30 \pi ^2\Big] \gamma ^2+
 216 c_1 \Big[6 c_2+24 c_3+(-25+12 \gamma_E) (\beta +4 \gamma )\Big]\\
+ 6 \ln\left(2 G_N M \mu \right) \Big[\beta  \left[432 c_1+36 (-63+40 \gamma_E) \alpha +(-181+168 \gamma_E) \beta \right]+192 c_3 (9 \alpha +\beta -2 \gamma )\\
+24 \left[72 c_1+48 (-8+5 \gamma_E) \alpha +(-11+20 \gamma_E) \beta \right] \gamma +64 (31-12 \gamma_E) \gamma ^2\\
+24 c_2 (18 \alpha +5 \beta +8 \gamma )+24 (36 \alpha +5 \beta -4 \gamma ) (\beta +4 \gamma ) \ln\left(2 G_N M \mu \right)\Big]\\
+\frac{2048 \pi ^4\rho Q^2}{9 G M^5} (4 \alpha +\beta )\bigg\{-36c_2^2+2880c_3^2+24c_3\Big[36(1+\gamma_E)\alpha + (-113+48\gamma_E)\beta\Big]\\
+\beta\Big[9\left[-71+24\gamma_E(-3+2\gamma_E)\right]\alpha -8\left[40+3\gamma_E(-23+6\gamma_E)+3\pi^2\right]\beta\Big]\\
+6c_2\Big[96c_3+36(1+\gamma_E)\alpha+2(23-12\gamma_E)\beta+(-223+192\gamma_E)\gamma\Big]\\
+12 \Big[2 c_3 (-859+480 \gamma_E)+3 \left[-71+24 \gamma_E (-3+2 \gamma_E)\right] \alpha \\+\left[695+3 \gamma_E (-225+64 \gamma_E)-18 \pi ^2\right] \beta \Big] \gamma
 +16 \Big[2405+3 \gamma_E (-859+240 \gamma_E)-30 \pi ^2\Big]\gamma ^2\\
+108 c_1 \Big[c_2+4 c_3+(-5+2 \gamma_E) (\beta +4 \gamma )\Big]+12 \ln\left(2 G_N M \mu \right) \Big[6 c_2 (3 \alpha -2 \beta +16 \gamma )\\
+24 c_3 (3 \alpha +4 \beta +40 \gamma )+(\beta +4 \gamma ) \big[18 c_1+18 (-3+4 \gamma_E) \alpha +2(23 -12 \gamma_E) \beta \\+(-859 +480 \gamma_E) \gamma \big]+12 (\beta +4 \gamma ) (3 \alpha -\beta +20 \gamma ) \ln\left(2 G_N M \mu \right)\Big]\bigg\}.
\end{gathered}
\end{equation}
\end{document}